\documentstyle[aps,twocolumn,floats,graphicx]{revtex}
\begin{document}
\title{{\bf Positron Tunnelling through the Coulomb Barrier of Superheavy Nuclei}}
\date{CUQM-77 \qquad SUSX-TH/00-002}
\author{{\bf Norman Dombey}$^{(a)}$ and {\bf Richard L. Hall}$^{(b)}$}
\address{$^{a}$Centre for Theoretical Physics, University of Sussex, Brighton BN1\\
9QJ,~UK\\
email: normand@sussex.ac.uk}
\address{$^{b}$Department of Mathematics and Statistics, Concordia University,\\
Montreal, Quebec, Canada H3G 1M8\\
email: rhall@cicma.concordia.ca}
\address{\flushleft\rm
We study beams of medium-energy electrons and positrons which obey the Dirac
equation and scatter from nuclei with $Z > 100.$ At small distances the
potential is modelled to be that of a charged sphere. A large peak is
found in the probability of positron penetration to the origin
for $Z \approx 184.$ This may be understood as an example of Klein
tunnelling through the Coulomb barrier: it is the analogue of the Klein
Paradox for the Coulomb potential. 
}
\maketitle
%%%%%%%%%%%%%%%%%%%%%%%%%%%%%%%%%%%%%%%%%%%%%%%%%%%%%%%%%%%%%%%%%%%%%
\section{Quantum Tunnelling and Klein Tunnelling}

One of the principal characteristics of quantum mechanics as opposed to
classical mechanics is that particles of energy $E$ can pass through regions
of a potential $V(r)$ in which their kinetic energy $T=E-V$ is negative,
albeit with amplitudes which are exponentially suppressed. In two pioneering
papers 70 years ago which demonstrated the importance of quantum tunnelling
in the understanding of nuclear decays, the probability of an $\alpha $%
-particle to tunnel from the potential well of the nucleus through the
Coulomb barrier was calculated \cite{GGC}. The calculation in the
quasi-classical case is straightforward\cite{LL}: the transition probability
for decay $w$ is given by

\begin{equation}
w\sim \exp \left\{ -2\int_0^{r_c}\sqrt{2m\left[ V(r)-E\right] }\right\}
=\exp \left\{ -2\pi Z\alpha m/p\right\}  \label{coul}
\end{equation}

\noindent  for $V(r)=Z\alpha /r$ where $r_c=Z\alpha /E$ is the classical
turning point, and $m$ and $p$ are the mass and momentum of the $\alpha $%
-particle. The integral in the exponent is over the classically forbidden
region in which $T<0$. Similarly a positron of energy $E,$ momentum $p,$ and
mass $m$ incident on a nucleus of charge $Z$ can reach distances smaller
than $r_c.$ If $\rho =\left| \psi (0)\right| _{pos}^2/\left| \psi (0)\right|
_{el}^2$ is the ratio of the probability of a positron penetrating a Coulomb
barrier to reach the origin compared with the probability of an electron of
the same energy, then essentially the same calculation gives
non-relativistically

\begin{equation}
\rho =e^{-2\pi y}  \label{coul2}
\end{equation}

\noindent  where $y=Z\alpha m/p$ so that $\rho $ decreases exponentially
with $Z$ for fixed $p$. If the particles are relativistic and satisfy the
Dirac equation then Eq.(\ref{coul2}) is still obtained provided now $%
y=Z\alpha E/p$ \cite{rose}. For relativistic problems the kinetic energy $%
T=E-V-m=\sqrt{m^2+p^2}-V-m$ so that, classically, $T$ is still positive
definite but now $r_c=Z\alpha /(E-m).$

At the same time as the papers showing the effect of quantum tunnelling in $%
\alpha $-decay, Klein \cite{klein} showed that electrons in the Dirac
equation could in principle tunnel through a high repulsive barrier. This is
the famous Klein Paradox. Calogeracos and one of us (ND) have recently
reassessed this phenomenon \cite{kleinf} and call such quantum tunnelling
without the expected exponential suppression {\it Klein tunnelling}. To
obtain Klein tunnelling it is essential that hole states (corresponding to
negative energy states for free particles) as well as particle states are
considered and allowed to propagate. For a particle of momentum $p$ and
energy $E$ incident on a Klein step of height $V$, a hole state of momentum $%
-q$ will propagate under the barrier provided 
\[
V=\sqrt{m^2+p^2}+\sqrt{m^2+q^2}, 
\]
\noindent which is possible for $V>2m.$ In terms of the particle kinetic
energy under the barrier 
\[
T=E-V-m=-m-\sqrt{m^2+q^2}\leq -2m, 
\]
\noindent  and where $T\leq -2m$ hole states can propagate without
exponential suppression. $T\leq -2m$ for a Coulomb potential corresponds 
to penetrating under the
barrier to distances $r<r_K,$ where $r_K=Z\alpha
/(E+m)<r_c$ is the Klein distance. While tunnelling from $r_c$ to $r_K$ is
exponentially suppressed, tunnelling from $r_K$ to $r=0$ is not; indeed the
amplitude may even be enhanced. This is because the effective potential in a
relativistic theory with a potential $V$ which is the time component of a
four-vector is given by

\begin{equation}
2mV_{eff}(r)=2EV(r)-V^2(r).  \label{square}
\end{equation}

\noindent  In the region $0<r<r_K,$ $V(r)>E+m>E$ and so the effective
Coulomb force is attractive \cite{schiff} when $r$ is small enough [note
that the force becomes attractive when $dV_{eff}/dr$ changes sign, not when $%
V_{eff}(r)$ changes sign].

When the Coulomb potential is investigated in the Dirac equation, the ground
state energy $E_1=0$ at $Z\alpha =1$ and becomes complex for $Z\alpha >1$ 
\cite{rose}. This is a consequence of the $V^2(r)$ term in Eq. (\ref{square}%
) which leads to the ``collapse'' of the particle to the origin, and hence
to a problem which is not well-defined \cite{case}. So the theory breaks
down at $Z_{\max }=1/\alpha \simeq 137$. Since superheavy nuclei can be
constructed in heavy ion collisions with values of $Z$ larger than $137$,
this limitation on $Z$ cannot be physical. A modified Coulomb potential
which takes account of the finite size of the nucleus must therefore be used
so that the singularity at small $r$ is smoothed out. When this is done
there seems to be no restriction on $Z$; furthermore bound state energies
become negative for sufficiently large $Z.$ These are interpreted as bound
positron states \cite{grein2}, \cite{zeld}. For Klein tunnelling through
the (modified) Coulomb potential it was conjectured \cite{kleinf} that it
might occur at values of $Z$ large enough to obtain negative energy bound
states. For the simplest such potential, due to Pieper and Greiner \cite
{grein2}, the first negative energy bound state occurs at $Z=147$. Hence the
prediction of Calogeracos and Dombey \cite{kleinf} that Klein tunnelling may
occur for nuclei with $Z\approx 150$ or above. If it occurs, Klein
tunnelling would lead to a breakdown of the exponential suppression given by
Eq.(\ref{coul2}).

In the Dirac equation in one-dimension, transmission resonances 
(with zero reflection coefficient) 
\cite{bohm} have been demonstrated for electron scattering off square
barriers \cite{grein} and off smooth potential barriers of Woods-Saxon \cite
{jens} or Gaussian \cite{ken} form. These provide examples of Klein
tunnelling. In each case, the transmission resonances occur when the
corresponding attractive potential becomes supercritical (see next
paragraph).

Since it is unlikely that superheavy nuclei with $Z>150$ can be prepared and
stay around for long enough for positrons to be scattered off them, we study
positron scattering off nuclei with $Z>150$ numerically by solving the Dirac
equation for a modified Coulomb potential. We thus attempt to repeat Klein's
analysis for a modified Coulomb potential in place of a potential step. We
do indeed find that positrons scattering off nuclei with $Z>150$ and
especially with $Z>170$ no longer satisfy Eq.(\ref{coul2}). We find, in
addition, an extremely sharp peak in $\rho $ near $Z=184$. We now outline
the calculation.

\section{Modified Coulomb Potential: Bound and Continuum states}

Following Pieper and Greiner\cite{grein2}, the modified Coulomb potential
experienced by an electron or positron beam is taken to be the potential
arising from a homogeneously charged sphere of radius $b$. More explicitly,
the time component $V(r)$ of the 4-vector potential is given by 
\[
V(r)=\pm {\frac{{Z\alpha }}{{b}}}f(r/b), 
\]
\noindent where the dimensionless potential shape function $f$ is defined as 
\[
f(r/b)=-\frac 32+\frac 12\left( \frac rb\right) ^2\quad r\leq b 
\]
\[
f(r/b)=-\frac br\quad r>b 
\]
\noindent and $b={{(1.2)A^{\frac 13\text{ }}}}$fermi. We let $\psi _1$ and $%
\psi _2$ be the `large and small' radial wave functions used to construct
the Dirac spinor corresponding to a total angular momentum of $j.$ We employ
the variables $\tau =\pm 1$, and $k=j+{\frac 12}$, so that the parity $P$ of
the spinor is given by $P=(-1)^{j+{\frac \tau 2}}=\pm 1$. In the notation of
Rose\cite{rose}, with $\kappa =\tau k,$ the coupled radial equations may be
written

\begin{equation}
\psi _2^{\prime }-{\frac{{\tau k}}{{r}}}\psi _2=(m+V-E)\psi _1  \label{dir1}
\end{equation}

\begin{equation}
\psi _1^{\prime }+{\frac{{\tau k}}{{r}}}\psi _1=(m-V+E)\psi _2.  \label{dir2}
\end{equation}
\noindent There are two distinct problems to be considered: (a) bound
states; and (b) continuum states. For the bound states, we adopt the
boundary conditions and normalization given by: 
\[
\psi _1(0)=\psi _2(0)=0,\quad \int_0^\infty (\psi _1^2(r)+\psi _2^2(r))dr=1. 
\]
\noindent We now choose $j$ and $\tau ,$ integrate out from the origin, and
search for those energies which lead to spinors that have the desired number 
$n_1$ of nodes in the `large' radial component $\psi _1,$ and vanish at
large distances. If the quantity $\ell =j+\frac 12\tau $ represents the
orbital angular-momentum quantum number in the first two components of the
Dirac spinor, and $\nu =(n_1+1+\ell ),$ then the usual spectroscopic
designation is written $\nu \ell _j,$ where $\ell =\{0,1,2,\dots \}\sim
\{s,\ p,\ d,\dots \}.$ The $j=\frac 12,\,l=0$ states, for example,
correspond to $\tau =-1$. We calculate the first few energy eigenvalues as a
function of $Z$ and our results agree with those given by Pieper and Greiner
in their Table~3. In particular, we have found the critical values of the
atomic number $Z$ that yield the energies $E=0$ (the threshold for bound
positrons) and the supercritical values of $Z$ corresponding to energy $E=-m$
(corresponding to spontaneous positron production)$;$ some results are
presented in Table~1

\vskip 0.1 true in

\begin{center}
\begin{tabular}{|c|c|c|}
\hline
& $E=0$ & $E=-m$ \\ \hline
$1s_{1/2}$ & $146.7$ & $170.4$ \\ \hline
$2s_{1/2}$ & $195.7$ & $237.0$ \\ \hline
$2p_{1/2}$ & $168.1$ & $183.8$ \\ \hline
$3s_{1/2}$ & $260.1$ & $316.5$ \\ \hline
\end{tabular}
\vskip 0.1 true in

{\em Table 1}: {\it Values of }$Z${\it \ for which }$E=0${\it \ and }$E=-m$
\end{center}

\noindent In the case of the continuum states, we adopt the following
method. At very small distances $r<<b,$ we assume that the large and small
radial functions have the asymptotic pure-power forms $Cr^\beta$ and the ratios
given (after some elementary corrections) by Rose\cite{rose}. In this region
there is then only one free parameter, which we take to be the amplitude $%
C_{1e},$ or, for positrons, $C_{1p},$ of the large component $\psi _1(r);$
the small component is smaller here by the factor $r^2.$ We fix $j,$ $\tau ,$
and the incident momentum $p,$ and we integrate outwards, to a point $r_b$ well
beyond the classical turning point $r_c$. At these distances the radial
components have the large-$r$ Coulombic asymptotic form, which is
approximately sinusoidal with $\psi _1$-amplitude, say $A_1;$ in this
region, the components are also (almost) exactly out of phase, so that when $%
\psi _1=A_1,$ we have $\psi _2=0.$ We now search, in each case, for the
value of $C_1$ so that this asymptotic amplitude has the value $A_1=1.$
Since the numerical process is, in principle, direction invariant, this
corresponds to an `experiment' in which the final amplitude of an incoming
beam with unit amplitude is determined near the origin. Since we are
calculating wave functions at the origin, only $l=0$ states contribute and
thus we restrict the analysis to $\tau =-1$ and $k=1$. For a given $Z$ we
can express the positron-electron ratio in the form $\rho
=[C_{1p}/C_{1e}]^2. $

\section{The Peak}

As we are not interested in the detailed behaviour of the nuclear charge
distribution we require $b<<r_K.$ We also require $r_c>>r_K$ so that the
normal exponential suppression will be obtained for $Z$ not too large. We
ensure that this is so by choosing the momentum $p$ in the range $0.1m-0.4m$.
For this momentum range and with $Z<150$ we expect normal exponential
suppression of the positrons by the modified Coulomb potential according to
Eq.(\ref{coul2}): this is demonstrated in Fig.~1 where our calculation of $%
\rho $ for $Z=100$ is close to $e^{-2\pi y}$ asymptotically, showing that
our method of calculation is satisfactory. 
%%%%%%%%%%%%%%%%%%%%%%%%%%%%%%%%%%%%%%%%%%%%%%%%%%%%%%%%%%%%%%%%%%%%%%%%%%%%%
\begin{figure}[th]
%h=here,t=top,b=bottom,p=separate figure page
\par
\begin{center}
\leavevmode
\includegraphics[width=0.7\linewidth]{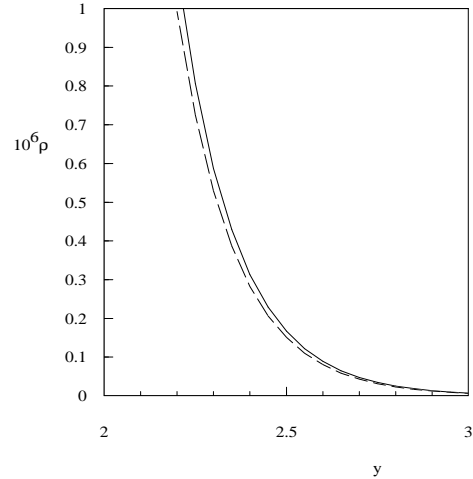} \medskip
\end{center}
\caption{A comparison of $\rho $ (solid) with $e^{-2\pi y}$ (dashed) for Z =
100.}
\end{figure}
%%%%%%%%%%%%%%%%%%%%%%%%%%%%%%%%%%%%%%%%%%%%%%%%%%%%%%%%%%%%%%%%%%%%%%%%%%%%%

We now look at $C_{1p}/C_{1e}$ for values of $Z>140$. The principal
discovery we report here is the existence of a pronounced peak in $\rho =
[C_{1p}/C_{1e}]^{2}$ very near $Z=184.$ This peak is exhibited in Fig.~2 for
the case $p=(0.4)m.$ At the peak maximum $\rho \approx 2$. The peak is so
sharp that it seems to correspond to a singularity of some sort. 
%%%%%%%%%%%%%%%%%%%%%%%%%%%%%%%%%%%%%%%%%%%%%%%%%%%%%%%%%%%%%%%%%%%%%%%%%%%%%
\begin{figure}[ht]
%h=here,t=top,b=bottom,p=separate figure page
\par
\begin{center}
\leavevmode
\includegraphics[width=0.7\linewidth]{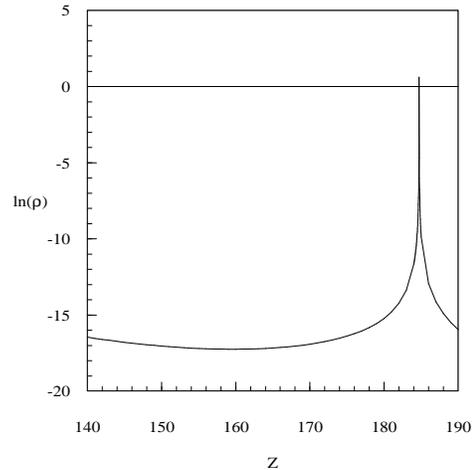} \medskip
\end{center}
\caption{$ln(\rho)$ as a function of $Z$}
\end{figure}
%%%%%%%%%%%%%%%%%%%%%%%%%%%%%%%%%%%%%%%%%%%%%%%%%%%%%%%%%%%%%%%%%%%%%%%%%%%%%
It is convenient to look at the deviation from exponential suppression by
introducing a Klein ''logarithmic form factor'' $R$ so that $%
\left
|C_{1p}/C_{1e}\right | = e^{R}e^{-\pi y},$ where $R$ is given
explicitly by

\begin{equation}
R(Z,p) = \ln \left |C_{1p}/C_{1e}\right |+\pi y = {\frac 12}\ln(\rho) + \pi
y.  \label{R}
\end{equation}

\noindent Thus $R$ constant corresponds to normal exponential suppression.
In Fig.~3 we exhibit more detail near the peak by plotting $R$ in terms of $%
Z $ for the two cases $p=0.1m$ and $0.4m.$ The peak shifts slightly with
momentum, with a limit, for $p\rightarrow 0,$ very close to the $p=0.1$
position shown. More specifically, we find peaks in $\rho (Z,p)$ at
positions $\hat Z_p$ given by $\hat Z_{0.4}=184.8$ and $\hat Z_{0.1}=\hat Z%
_{0.02}=183.8.$ This is fully consistent with a peak in $\rho (Z,p=0)$ at $%
\hat Z_0=$ $183.8.$%
%%%%%%%%%%%%%%%%%%%%%%%%%%%%%%%%%%%%%%%%%%%%%%%%%%%%%%%%%%%%%%%%%%%%%%%%%%%%%
\begin{figure}[tbph]
%h=here,t=top,b=bottom,p=separate figure page
\par
\begin{center}
\leavevmode
\includegraphics[width=0.7\linewidth]{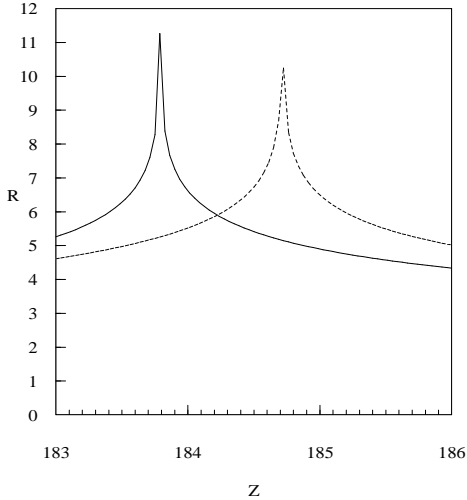} \medskip
\end{center}
\caption{$R$ as a function of $Z$ for $p=0.4m$ (dashed) and $p=0.1m$ (solid)}
\end{figure}
%%%%%%%%%%%%%%%%%%%%%%%%%%%%%%%%%%%%%%%%%%%%%%%%%%%%%%%%%%%%%%%%%%%%%%%%%%%%%

\section{Discussion}

The enormous peak near $Z=184$ in the scattering of positrons by superheavy
nuclei was unexpected: it involves an enhancement in $\rho $ by a factor $%
10^8$ between $Z=179$ and $Z=184.$ While a demonstration of a deviation from
the pure exponential suppression of Eq.(\ref{coul2}) resulting from Klein
tunnelling was the purpose of this calculation, we initially expected it to
be a simple maximum in $\rho $ at the first supercritical value of $Z=170.4$
as is the case with the one-dimensional examples referred to above, where
supercriticality and the maximum of the transmission coefficient coincide.
While the peak that we have found does not correspond to the first $1s_{1/2}$
supercritical state, it does correspond to the second $2p_{1/2}$
supercritical state (see Table $1$). We show in a forthcoming
paper \cite{new} that this correspondence exists independently of the exact
form of the potential. This is also in agreement with the result of
M\"uller, Rafelski and Greiner \cite{grein3} who predicted 
a resonance in the negative energy continuum 
in the $s_{1/2}$ state at $Z=184$.

To explain why the peak occurs in the state which corresponds to the $%
2p_{1/2}$ bound state of the electron is an exercise in Dirac's hole theory.
The vacuum state for our modified Coulomb potential at supercriticality
contains a vacant $2p_{1/2}$ bound state of an electron of energy $E=-m;$
this in hole theory represents a positron state of zero kinetic energy. But
to compare the original electron state with a positron state
requires charge conjugation: equations (\ref{dir1},\ref{dir2}) are invariant
under the combined operation $E\rightarrow -E;V\rightarrow -V;\tau
\rightarrow -\tau ;\psi _1\Leftrightarrow \psi _2.$ So the vacant electron $%
2p_{1/2}$ bound state corresponds to a resonant $s_{1/2}$ positron scattering state
since $j=\frac 12$ for both states but $\tau $ changes sign.\footnote{%
We thank Dr. X. Artru for pointing this out.}

The (vacant) electron $2p_{1/2}$ bound state of zero kinetic energy at
supercriticality can thus be considered as a positron resonance of zero
kinetic energy. In one dimension non-relativistically a zero energy resonance
is a transmission resonance where the transmission coefficient is unity and
the reflection coefficient is zero \cite{senn}. It seems likely that we have
found a similar effect here in three dimensions since our peak occurs as
close to zero kinetic energy (corresponding to the supercritical energy) as
we can calculate. We should emphasise that we calculate the quantity $\rho $
which is not a scattering cross section. Neither is it strictly a
transmission coefficient.  Nevertheless our peak is more like a transmission
resonance than a resonant cross section.  We will examine this point in more
detail in a forthcoming paper.

We note finally that our example here of positron scattering by a modified
Coulomb potential shows that the phenomenon discovered by Klein of Dirac
particles tunnelling through repulsive potentials is a general
characteristic of the Dirac equation in the presence of strong fields.

\noindent {\it Acknowledgments}~~One of us (RLH) would like to thank the
Natural Sciences and Engineering Research Council of Canada for partial
support of this work under grant No. GP3438. The other (ND) thanks Gabriel
Barton and Alex Calogeracos for their help.

\end{document}